\documentclass[]{interact}
\usepackage[numbers,sort&compress,merge]{natbib}% Citation support using natbib.sty
\bibpunct[, ]{[}{]}{,}{n}{,}{,}% Citation support using natbib.sty
\renewcommand\bibfont{\fontsize{10}{12}\selectfont}% Bibliography support using natbib.sty

\theoremstyle{plain}% Theorem-like structures provided by amsthm.sty

\theoremstyle{definition}

\theoremstyle{remark}

\usepackage{verbatim}

\begin{document}

\title{New measurement of the diffusion of carbon dioxide on non-porous amorphous solid water}

\author{
\name{Jiao He\textsuperscript{a}\thanks{Corresponding author Jiao He, Email: he@mpia.de}, Paula Caroline P\'erez Rickert\textsuperscript{a,b}, Tushar Suhasaria\textsuperscript{a}, Orianne Sohier\textsuperscript{a,c}, Tia B{\"a}cker\textsuperscript{a,d}, Dimitra Demertzi\textsuperscript{a,e}, Gianfranco Vidali\textsuperscript{f} and Thomas K. Henning\textsuperscript{a}}
\affil{
\textsuperscript{a}Max Planck Institute for Astronomy, K{\"o}nigstuhl 17, D-69117 Heidelberg, Germany; 
\textsuperscript{b}Department for Physics and Astronomy, Heidelberg University, 69120 Heidelberg, Germany; 
\textsuperscript{c}Département de chimie, École normale supérieure, PSL University, 75005 Paris, France;
\textsuperscript{d}Physics Department, University of Helsinki, Helsinki, 00560, Finland;
\textsuperscript{e}Institute for Molecules and Materials, Radboud University, Nijmegen 6525 AJ, Netherlands; 
\textsuperscript{f}Physics Department, Syracuse University, Syracuse, NY 13244, USA;
}
} 

\maketitle

\begin{abstract}
The diffusion of molecules on interstellar grain surfaces is one of the most important driving forces for the molecular complexity in the interstellar medium. Due to the lack of laboratory measurements, astrochemical modeling of grain surface processes usually assumes a constant ratio between the diffusion energy barrier and the desorption energy. This over-simplification inevitably causes large uncertainty in model predictions. We present a new measurement of the diffusion of CO$_2$ molecules on the surface of non-porous amorphous solid water (np-ASW), an analog of the ice mantle that covers cosmic dust grains. A small coverage of CO$_2$ was deposited onto an np-ASW surface at 40~K, the subsequent warming of the ice activated the diffusion of CO$_2$ molecules, and a transition from isolated CO$_2$ to CO$_2$ clusters was seen in the infrared spectra. To obtain the diffusion energy barrier and pre-exponential factor simultaneously, a set of isothermal experiments were carried out. The values for the diffusion energy barrier and pre-exponential factor were found to be $1300\pm110$~K and $10^{7.6\pm0.8}$~s$^{-1}$. A comparison with prior laboratory measurements on diffusion is discussed. 
\end{abstract}


\section{Introduction}
\label{sec:intro}
Dust grains in molecular clouds are covered by an ice mantle consisting of H$_2$O, CO$_2$, CO, NH$_3$, CH$_4$, and CH$_3$OH\cite{Boogert2015}. Other molecules, particularly complex organic molecules, which are defined in the astronomy context as carbon-containing molecules with at least six atoms, should also be present in the ice\cite{Herbst2009}. Various laboratory experiments and astrochemical modeling suggest that most of these molecules are formed by chemical reactions in the solid state rather than formed in the gas phase and then condensed on the grain surface\cite{Herbst2009}. After their formation, they may desorb, either thermally or non-thermally, and return to the gas phase and be identified by radio telescopes. 

One of the most important mechanisms for chemical reactions on grain surfaces is diffusion. Reactants on the surface need to diffuse in order to encounter the reaction counterparts. There are two types of diffusion, quantum diffusion, which is only important for very light species such as hydrogen, and thermal diffusion, which is relevant for almost all reactants. In most gas-grain astrochemical models, thermal diffusion is the dominant pathway to chemical reactions on grains\cite{Garrod2006, Acharyya2022}, although in recent years the role of thermal diffusion has been challenged and non-diffusive mechanisms have been proposed as well to account for the formation of complex organic molecules in interstellar ices\cite{Jin2020, Garrod2022, Herbst2022, He2021phase, He2022phase}.

Thermal diffusion is normally described by an Arrhenius equation with two parameters, the pre-exponential factor $\nu_{\rm dif}$, and the diffusion energy barrier $E_{\rm dif}$. The goal of laboratory experiments on diffusion is to quantify $E_{\rm dif}$ and $\nu_{\rm dif}$ for various atoms, radicals, and molecules on representative dust grain surfaces, such as water ice, silicate, or carbonaceous surfaces. Diffusion of reactants on silicate and carbonaceous surfaces is only important before the dust grains are covered by an ice layer. Once water starts to form on the bare grain surface, subsequent condensation of particles happens on top of the ice surface dominated by H$_2$O. Most of the existing laboratory experiments on diffusion focused on amorphous water ice, i.e., the so-called amorphous solid water (ASW), which is the dominant form of water ice on the grain surface. This current study also focuses on the diffusion on ASW surface. 

Several laboratory experimental studies focused on the diffusion of molecules on the surface of ASW, either on the outer surface of non-porous ASW (np-ASW) or through the inner pores of porous ASW (p-ASW). Zubkov et al. \cite{Zubkov2007} deposited N$_2$ on ASW of different thicknesses and performed temperature-programmed desorption (TPD) experiments, from which they obtained $E_{\rm dif}=890$~K and $\nu_{\rm dif}=9\times10^7$--$4\times10^8$~s$^{-1}$, respectively. Mispelaer et al. \cite{Mispelaer2013} presented one of the first systematic experimental studies on the diffusion of astro-relevant molecules in p-ASW. They covered the target molecule with ASW layers, and then warmed up the ice to a temperature at which the target molecules start to penetrate through the p-ASW and desorb from the ice. By monitoring the remaining amount of target molecules in the ice, they were able to quantify the diffusion rate through the pores. Though it is an important step ahead, it still has some limitations: (1) The diffusion might have already started even during the growth of the ASW layers, and it is not accurate to treat the beginning of the isotherm as the beginning of the diffusion process; (2) The structure of the ASW might be changing during the diffusion/desorption, since no annealing was done to stabilize the structure; (3) The effect of diffusion and desorption are mixed and very difficult to separate, therefore introducing large uncertainty; (4) Based on the design of the experiment, only a narrow temperature range can be explored, and the accuracy in diffusion energy barrier determination is limited. More recently, Mat{\'e} et al. \cite{Mate2020} used a similar approach to measure the diffusion rate of CH$_4$ in ASW, and therefore inherited the same advantages and drawbacks. 

Lauck et al. \cite{Lauck2015} measured the diffusion of CO in p-ASW. In their experiments, CO and water were deposited at 12~K. Then the ice was flash heated to a temperature between 15 and 23~K for isothermal experiments. They used the fact that pure CO and CO mixed with water ice has different infrared absorption peak shapes to trace the diffusion of CO in porous water ice. Pure CO ice has a sharp peak centered at about 2139~cm$^{-1}$, while the mixture of CO and ASW has two peaks at 2139 and 2152~cm$^{-1}$, respectively. As diffusion goes on, there is a gradual transition from pure CO to a mixture of CO and water. Therefore, the diffusion rate can be calculated from the time evolution of this transition. Compared to Mispelaer et al., Lauck et al. solved the problem of a limited temperature span, and the problem with desorption. However, the method by Lauck et al. suffers from an unstable ice structure during the diffusion process, and this method only works for CO diffusion on ASW. 

He et al. (2017) \cite{He2017co2diff} (referred to as He17 later) measured the diffusion of CO$_2$ on np-ASW. Discrete CO$_2$ molecules and CO$_2$ clusters have different infrared absorption peaks. With less than a monolayer of CO$_2$ present on the surface, when the surface temperature increases, CO$_2$ diffuses and forms clusters. Therefore, there is a transition from discrete CO$_2$ to CO$_2$ clusters which is observable in infrared measurement. By using a rate equation model to simulate the diffusion process and to fit the experimental data, the diffusion energy barrier was calculated to be 2100~K. It has to be noted that in this study, the prefactor was assumed to be $10^{12}$~s$^{-1}$, as is customarily done in these studies. If a different prefactor is assumed, the diffusion energy value would be different. It is important to determine both the prefactor and the diffusion energy value simultaneously. An energy value without the correct prefactor could be in error. 

Kouchi et al. (2020) \cite{Kouchi2020} used a transmission electron microscope (TEM) to measure the diffusion energy barrier of CO and CO$_2$ on p-ASW. This is the first attempt to use the TEM technique to measure the diffusion of astro-relevant molecules. In their experiment, they first deposited 10~nm of water ice at 10~K and then deposited, for example CO, while taking TEM scans during deposition. When the ice is covered by the required amount of CO, and if the temperature is high enough for the crystallization to proceed\cite{He2021phase}, crystals of CO are formed in the ice. They interpreted the forming of crystals as a result of the diffusion of CO and used the distance between crystals to represent the average diffusion distance of CO molecules. By analyzing the diffusion distance at different temperatures, they calculated the diffusion energy barrier. The study by Kouchi et al. suffers from a few drawbacks: (1) The interpretation of the experimental result is not intuitive, and alternative interpretations exist. The observed temperature dependence of crystal distance may be a result of the temperature dependence of the phase transition of CO ice. Such alternative interpretations have to be checked before extending this method to a wider range of systems; (2) The water ice was unstable during the diffusion and could have affected the experimental results; (3) They could not determine the prefactor from the experiments; as mentioned above, it is important to determine both values simultaneously. A similar approach has recently been applied to a wider range of molecules\cite{Furuya2022}. 

He et al. (2018)\cite{He2018diff} (referred to as He18 later) designed an experiment to measure diffusion based on the fact that when a molecule is binding to the dangling OH (dOH) site of p-ASW, the weak dOH absorption band in the infrared is red-shifted. In the experiment, porous water ice was grown at 10~K, stabilized by annealing to 70~K, and then cooled down to 10~K to deposit the target molecule. Then the ice was warmed to a specific temperature, for example, 20~K, and infrared spectra were measured continuously to monitor the shifting of dOH bands. When diffusion happens, a gradient in concentration of the target molecule is expected in the ice. After a long time, the whole pore surface is evenly covered by the target molecule, and almost all the dOH is at the shifted position. This method was applied to molecules including N$_2$, O$_2$, CO, CH$_4$, and Ar, all of which wet the ASW surface. From the experiments, both the prefactor and energy barrier for diffusion were obtained. The prefactor was found to be mostly in the $10^7$--$10^9$~s$^{-1}$ range, and the diffusion energy barrier was roughly 0.3 times the upper bound of the desorption energy distribution on ASW. Table~\ref{tab} shows a summary of the above-mentioned experimental studies and some key aspects of the experiments. The same method cannot be applied directly to non-wetting molecules such as CO$_2$ \cite{Noble2012}. In fact, we attempted to measure the diffusion of CO$_2$ through p-ASW using the same method, but found no clear evidence of CO$_2$ penetration into the p-ASW to form monomers. This is somehow in agreement with prior experimental studies on the segregation of CO$_2$ from CO$_2$:H$_2$O ice mixtures. Several groups found that by warming up the mixture of CO$_2$:H$_2$O ice, CO$_2$ tends to segregate and form clusters or islands, rather than mixing. This indicates that the interaction force between CO$_2$ molecules is stronger than the interaction force between CO$_2$ and ASW\cite{Hodyss2008, Oberg2009, He2018co2}. A TPD study of CO$_2$ on ASW also supports this argument\cite{He2017co2diff}. 

\begin{table}[ht]
  \caption{Summary of a few previous laboratory studies on diffusion. Mi13, La15, He17, He18, Ko20, Ma20 and Fu22 stand for Mispelaer et al. (2013)\cite{Mispelaer2013}, Lauck et al. (2015)\cite{Lauck2015}, He et al. (2017)\cite{He2017co2diff}, He et al. (2018)\cite{He2018diff}, Kouchi et al. (2020)\cite{Kouchi2020}, Mate et al. (2020)\cite{Mate2020}, and Furuya et al. (2022) \cite{Furuya2022}, respectively. The following aspects are considered: (1) whether desorption interferes with diffusion; (2) whether the water ice is stable during the whole diffusion process; (3) whether it is applicable to IR-inactive molecules; (4) whether both $E_{\rm dif}$ and $\nu_{\rm dif}$ are obtained simultaneously.  }
\begin{tabular}{l|cccccc}
\hline
                                     & Mi13           & La15          & He17                & He18         & Ko20/Fu22                 & Ma20     \\ \hline
Des. not interfere with dif. & $\times$      & $\checkmark$ & $\checkmark$          & $\checkmark$ & $\checkmark$          & $\times$     \\
H$_2$O ice stable during dif.    & $\times$      & $\times$     & $\checkmark$          & $\checkmark$ & $\times$              & $\times$     \\
Applicable to IR-inactive mol. & $\times$      & $\times$      & $\times$              & $\checkmark$ & $\checkmark$          & $\times$     \\
Both $E_{\rm dif}$ and $\nu_{\rm dif}$ obtained         & $\checkmark$  & $\checkmark$ & $\times$              & $\checkmark$ & $\times$              & $\checkmark$ \\ \hline
\end{tabular}\label{tab}
\end{table}

The mixed ice of CO$_2$:H$_2$O had been the focus of a number of experimental studies because of the presence of such mixture in interstellar ices\cite{Gerakines1995, Galvez2007, Galvez2008, Hodyss2008, Oberg2009, Fayolle2011}. However, diffusion of CO$_2$ on ASW surface was not addressed in those studies. Combining the strengths of both He17 and He18, we perform a new set of experiments to measure the diffusion rate of CO$_2$ on np-ASW. Isothermal experiments are carried out at different temperatures and CO$_2$ are allowed to diffuse and form clusters. We obtain the diffusion energy barrier $E_{\rm dif}$ and prefactor $\nu_{\rm dif}$ values simultaneously from the experiments. 

\section{Experimental}
\label{sec:exp}
\subsection{Experimental setup}
Experiments were performed using an ultra-high vacuum (UHV) setup, which has been described previously \cite{He2017co2diff, He2018diff, He2018co2}. The setup consists of a main UHV chamber and two molecular beamlines. In the present study, only the main UHV chamber was used. The chamber is pumped by a combination of scroll pumps, turbomolecular pumps, and a cryopump, and a base pressure of $2\times10^{-10}$~mbar is achieved after bake-out. Located at the center of the chamber is a gold-plated copper disk, which is used as the substrate onto which ice samples are grown. The substrate can be cooled down to 6~K using a closed-cycle helium cryostat, and heated up using a resistive heater. The temperature was measured using a silicon diode temperature sensor. A Lakeshore 336 temperature controller recorded and controlled the temperature to an accuracy of 0.05~K. Ice was grown on the substrate by gas/vapor deposition from the chamber background, following a procedure described in detail in He et al. (2018) \cite{He2018co2}. Two separate computer-controlled leak valves (VAT variable leak valve 590) were used to deposit H$_2$O and CO$_2$, respectively. The thickness of the ice in monolayer (ML, defined as 10$^{15}$ molecules per cm$^2$) was calculated by integrating the chamber pressure during deposition, assuming the sticking as unity \cite{he2016sticking}. The ionization pressure gauge calibration factors for H$_2$O and CO$_2$ were taken into account in the calculation. Throughout the experiments, a Thermo Fisher Nicolet 6700 Fourier Transform Infrared Spectrometer (FTIR) in the Reflection Absorption Infrared Spectroscopy (RAIRS) configuration was used to monitor the ice composition over time. A spectrum was created by taking an average of nine scans every 12 seconds.

\subsection{Experimental procedure}
\begin{figure}[ht]
  \centering
  \includegraphics[width=0.9\textwidth]{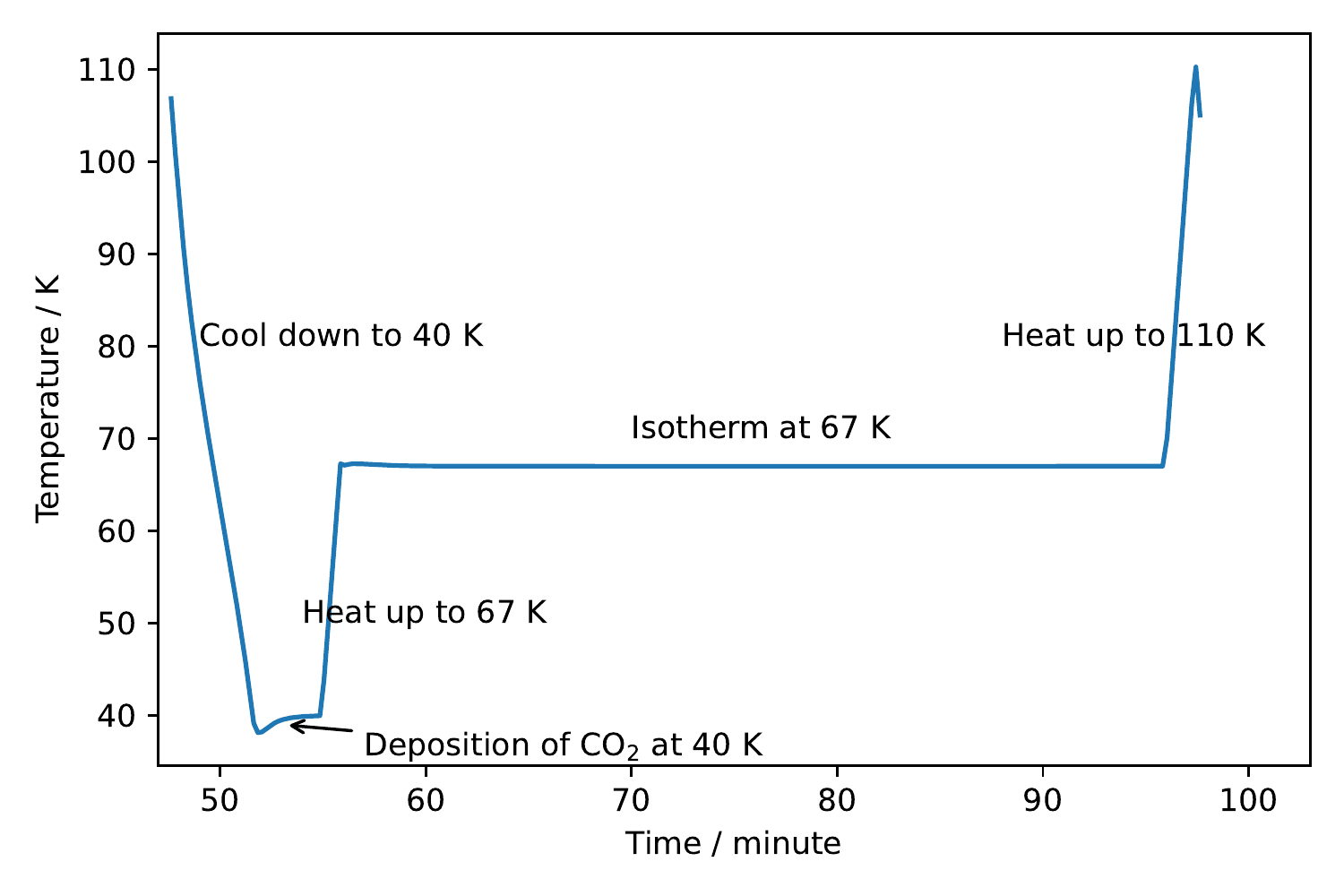}
  \caption{Temperature of the substrate during one segment of the experiment (isothermal experiment at 67~K). Different stages of the experiment are marked. } \label{fig:T_zoomin}
\end{figure}

The gold surface was first covered with 30 ML of non-porous amorphous solid water (np-ASW) by vapor deposition at a rate of 6 ML/minute over 5 minutes while the substrate was at 130~K. It remained at 130~K for 20 minutes to stabilize the ice and pump out the residual gas. Subsequently, the substrate was cooled down to 40~K for CO$_2$ deposition. 0.075 ML of CO$_2$ was deposited on top of the np-ASW at a deposition rate of 0.1 ML/minute. At 40~K, CO$_2$ does not diffuse\cite{He2017co2diff}, and the coverage is low enough so that almost all the CO$_2$ molecules are monomers. After CO$_2$ deposition, flash heating was applied to heat the substrate to the desired target temperature ($T_{\rm iso}$) at a ramp rate of 30~K/minute for the isothermal experiment. During the isothermal evolution, CO$_2$ molecules diffuse and form clusters. For $T_{\rm iso}>62$~K, the temperature was held for 30 minutes to allow CO$_2$ to diffuse and form clusters. For $T_{\rm iso}\leq62$~K, at which the diffusion is slower, 60 minutes were selected to provide sufficient time for diffusion. After the isothermal measurement, the sample was heated to 110~K to remove the CO$_2$ ice and leave the np-ASW intact. By 110~K, there is no CO$_2$ left on the surface, as evidenced by no signature of CO$_2$ in the RAIRS spectra. The substrate was then cooled down to 40~K again for the next CO$_2$ deposition. Figure~\ref{fig:T_zoomin} shows the temperature during one segment of the experiment. Different stages of the experiment are marked. To minimize human error, we used a LabVIEW program to regulate the temperature and gas deposition throughout the entire experiment. We explored the $T_{\rm iso}$ range between 56 and 68~K in 1~K steps.
%, and the temperature during the whole experiment is depicted in Figure~\ref{fig:T}.  

\section{Results and analysis}
\label{sec:results}
\begin{figure}[ht]
  \centering
  \includegraphics[width=0.9\textwidth]{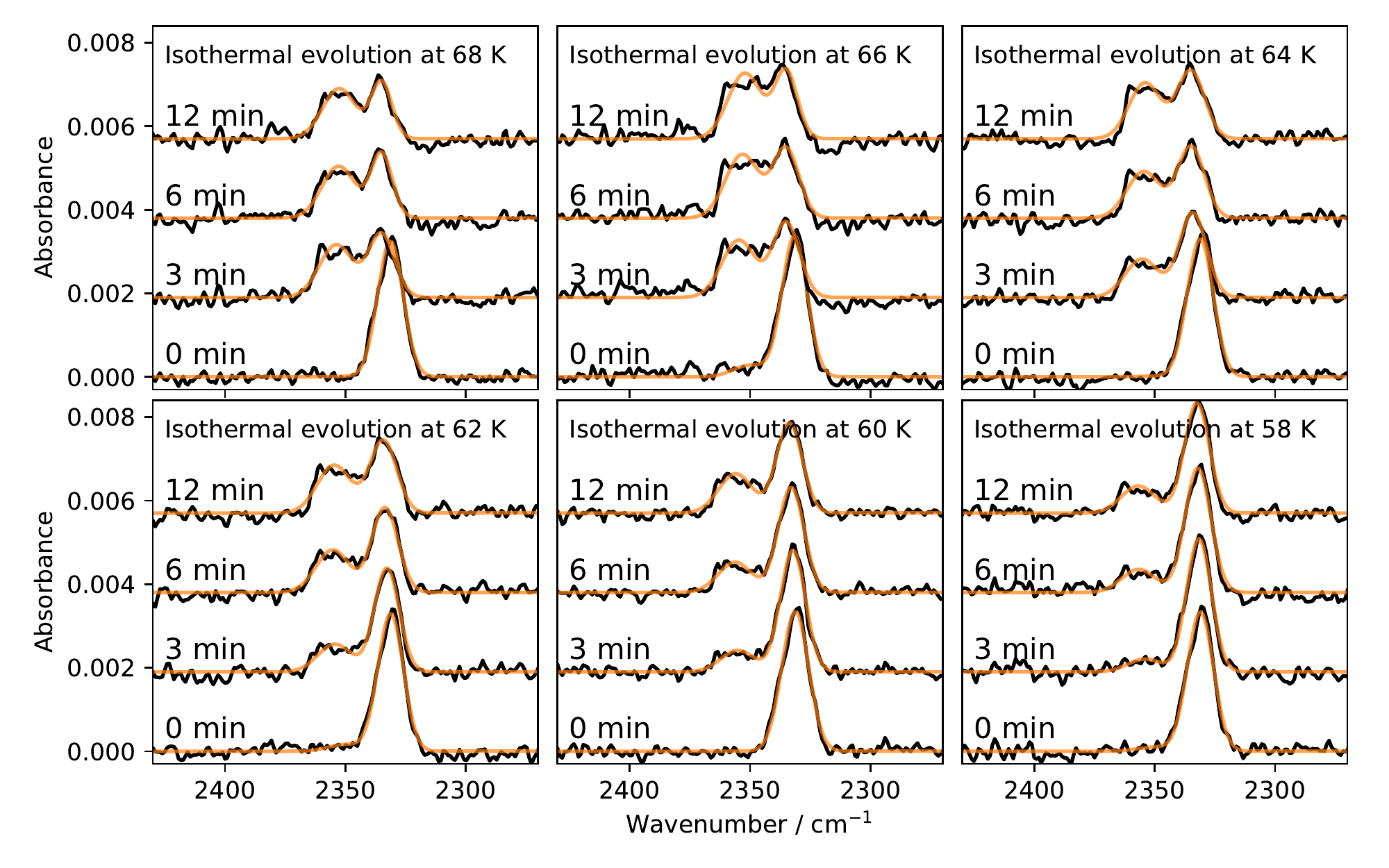}
  \caption{Selected RAIRS spectra in the range of the CO$_2$ $\nu_3$ mode during the isothermal experiment at different temperatures. Each panel depicts the CO$_2$ absorption peak at various times since the start of an isothermal experiment. The time and temperature of the isotherm are indicated in the figure. The black curves represent the experimental data, while the orange curves represent the fittings made with a sum of two Gaussian functions.} \label{fig:fitting}
\end{figure}

\begin{figure}[ht]
  \centering
  \includegraphics[width=0.5\textwidth]{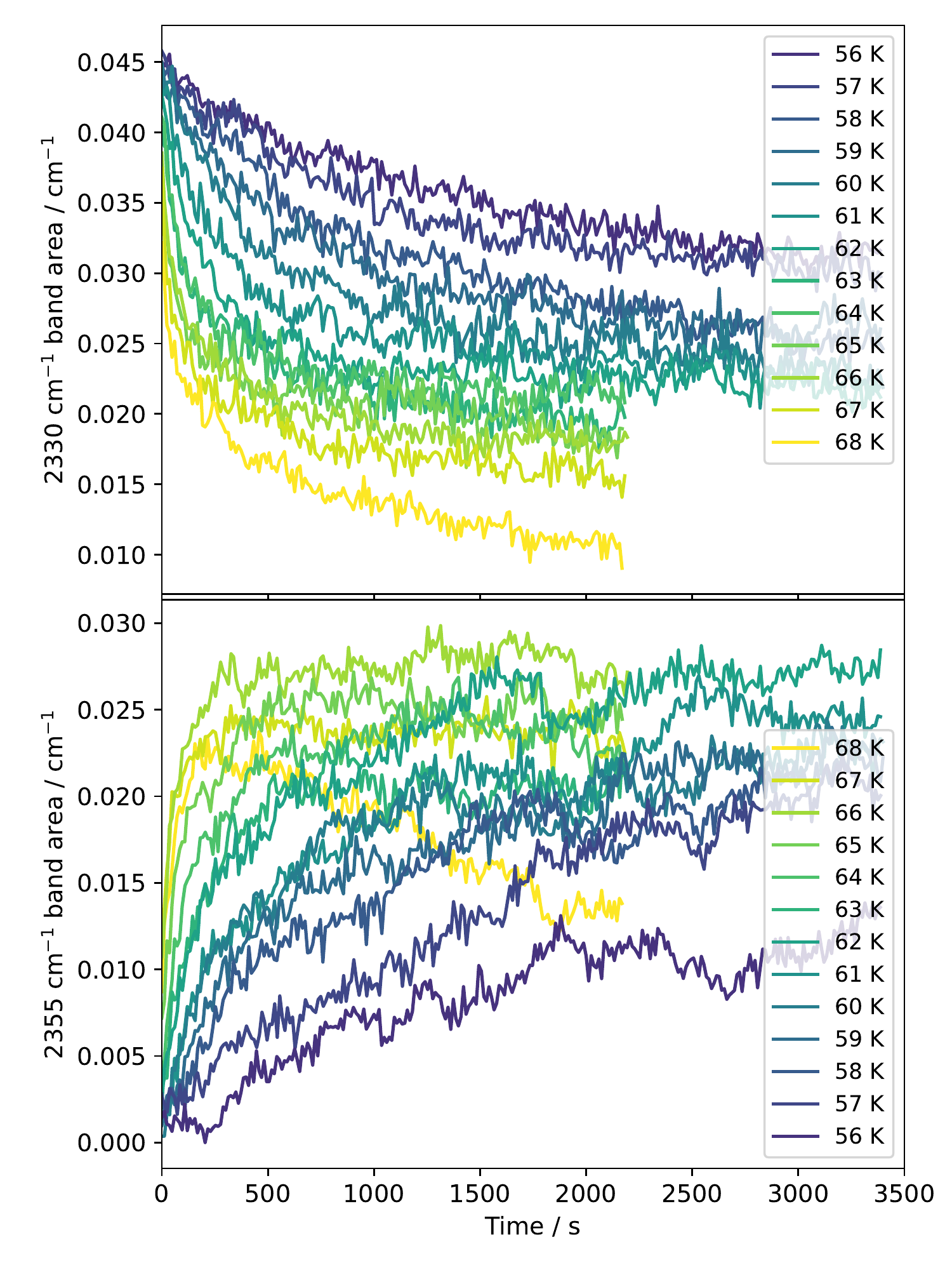}
  \caption{The band areas of the 2330 and 2355~cm$^{-1}$ components of the CO$_2$ absorption peak during the isothermal evolution at different temperatures.} \label{fig:bandArea}
\end{figure}

During the isothermal evolution at each temperature, RAIRS spectra are collected every 12 seconds, and a selection of them is displayed in Figure~\ref{fig:fitting}. At the very beginning of the isothermal evolution, the CO$_2$ absorption profile shows a single peak at 2330~cm$^{-1}$, which indicates isolated CO$_2$ molecules, i.e., monomers, on the surface. This is to be expected given that the coverage is only 0.075~ML and the likelihood of cluster formation is very low before the diffusion process kicks in. After a certain period of time, the 2330~cm$^{-1}$ peak drops and a new peak at $\sim$2355~cm$^{-1}$, indicative of CO$_2$ clusters \cite{He2017co2diff}, grows over time. This suggests that CO$_2$ is diffusing to form clusters. Considering the very low coverage of CO$_2$, we assume that the shifting of the adsorption peak position and the formation of clusters are only governed by the diffusion of CO$_2$ molecules.  The rate of diffusion should be temperature-dependent. Indeed, the shifting from 2330 to 2355~cm$^{-1}$ peak is faster at 68~K than at 58~K. To quantify the rate of diffusion, we use two Gaussian functions to fit the two peaks and analyze the time and temperature dependence of them. Figure~\ref{fig:fitting} shows the fitting of selected spectra using the sum of two Gaussian functions. This fitting is only an approximation, particularly, since the profile of the 2355~cm$^{-1}$ peak is not exactly a Gaussian line shape \cite{He2018co2}. Furthermore, even if all the CO$_2$ are in the form of clusters or in the form of bulk CO$_2$, the 2330~cm$^{-1}$ component is not absent. This is seen in a prior paper \cite{He2017co2diff}, which shows that even if CO$_2$ is allowed to have sufficient diffusion, both peaks are still present. Nonetheless, the shifting from one peak to the other still allows us to derive information about the kinetics. Figure~\ref{fig:bandArea} shows the resulting area of the two peaks during the isothermal evolution at each $T_{\rm iso}$. When the 2330~cm$^{-1}$ peak area decreases, the 2355~cm$^{-1}$ peak area increases, except when the temperature is higher than 67~K, at which the 2355~cm$^{-1}$ peak also drops. This is because the desorption of CO$_2$ starts when the temperature is close to $\sim$67~K \cite{He2017co2diff}, and the total area of the two peaks decreases over time. In the other extreme, when the temperature is at 56~K and the diffusion is too slow, the area of the 2355~cm$^{-1}$ peak is much noisier than the rest and is considered less reliable. As a result, for the diffusion analysis, we only consider the temperature range of 57 to 65~K.

\begin{figure}[ht]
  \centering
  \includegraphics[width=0.9\textwidth]{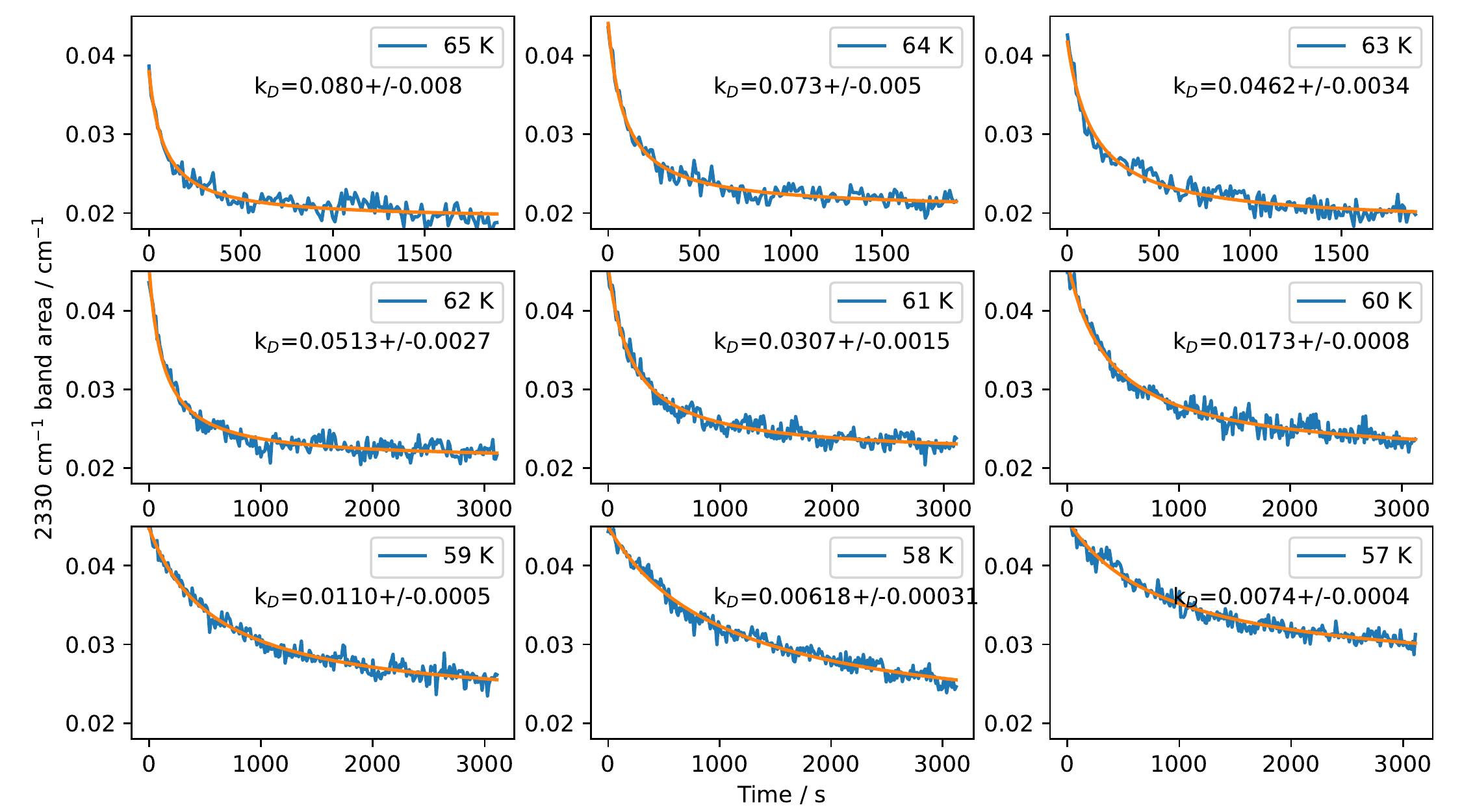}
  \caption{The band area of the 2330~cm$^{-1}$ component during isothermal experiment at different temperatures and the fitting using Equation~\ref{eq:c1_sol_linear}. The best fit $k_D$ values are shown in the figure. } \label{fig:Dfit}
\end{figure}

\begin{figure}[ht]
  \centering
  \includegraphics[width=0.9\textwidth]{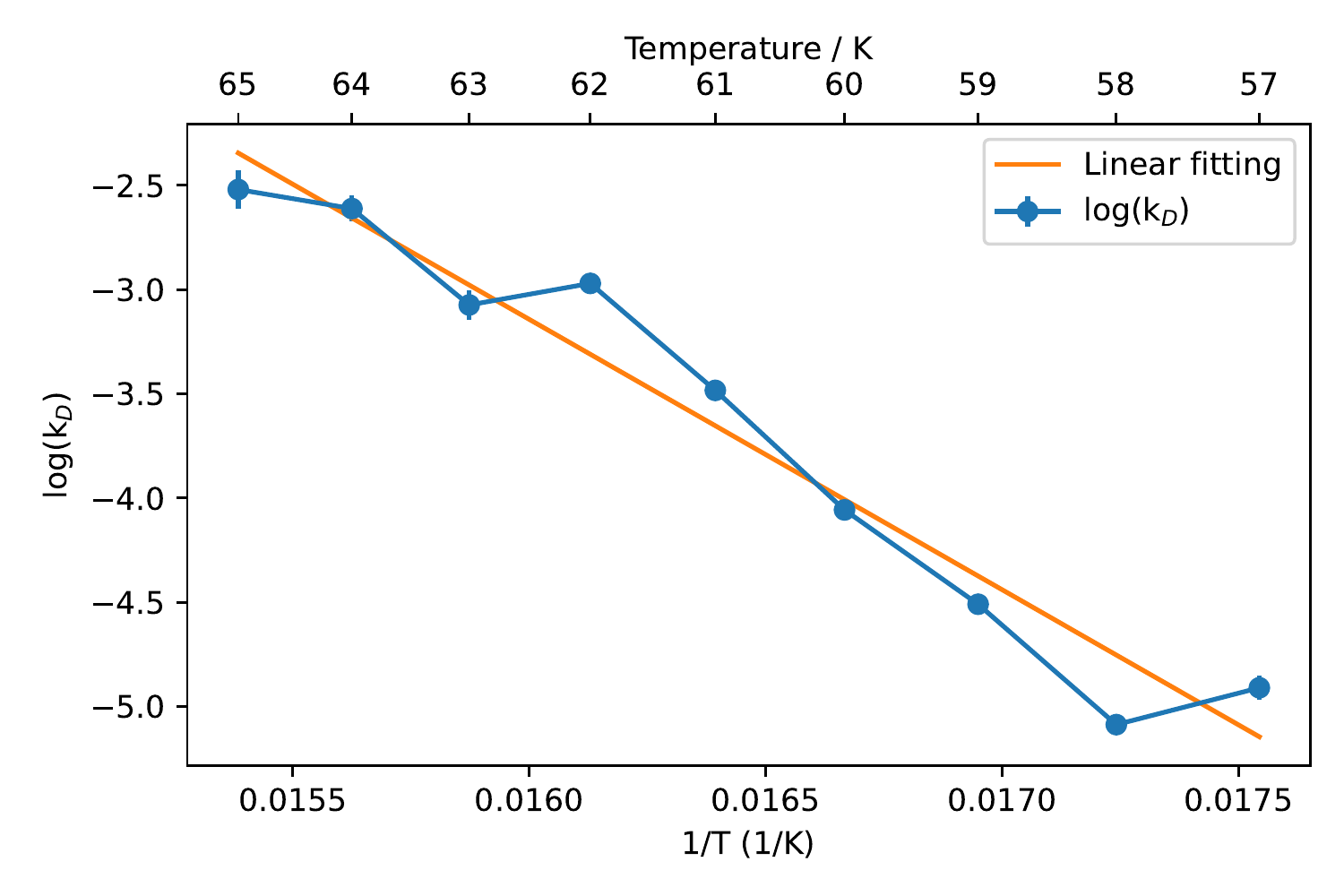}
  \caption{Arrhenius plot $log(k_D)$ over $1/T$, where $k_D$ is the diffusion rate, and $T$ is the isotherm temperature. The linear fitting is shown in orange. } \label{fig:arrh}
\end{figure}

Next, we use a model to describe the diffusion. Surface diffusion has been reviewed in previous publications \cite{Tsong2005,SurfaceDiffusion}. We refer intereted readers to those publications. When the concentration of molecules on the substrate surface is low enough so that the interaction force between them is negligible, the diffusion of molecules can be analyzed by assuming that they follow random walk. This type of diffusion is called tracer diffusion or intrinsic diffusion. On a 2D surface such as the np-ASW in the present study, CO$_2$ molecules jumps from one adsorption site to the nearest neighbour. Although the np-ASW surface is by no means a regular periodic surface, for the sake of simplicity, we still use a 2D lattice to represent it. We have assumed that 1 ML of CO$_2$ is equivalent to $10^{15}$ molecules per cm$^{2}$; correspondingly, the lattice size is $l=3.2$~\AA. In the random walk analysis, diffusion rate $k_D(T)$ can be described by an Arrhenius equation:
\begin{equation}
	k_D(T) = \nu_{\rm dif} \exp(\frac{-E_{\rm dif}}{k_{\rm B}T}) \label{eq:arr}
\end{equation}
$E_{\rm dif}$ is the diffusion energy barrier, and $\nu_{\rm dif}$ is the prefactor for diffusion. In the literature, the prefactor is sometimes referred to as the attempt frequency. In astronomy, it is a convention to omit the Boltzmann constant $k_{\rm B}$, and then the diffusion energy barrier has the unit Kelvin. We denote the time-dependent coverage of CO$_2$ monomers and dimers as $C_1(t)$ and $C_2(t)$, respectively. They are unitless variables defined to be the fraction of absorption sites holding monomers and dimers. In theory, larger clusters such as trimers should also be considered. However, due to the low coverage of CO$_2$, the coverage of trimers should be much smaller than dimers and can be ignored. Once dimers are formed, we assume that they do not break apart before desorption. This is reasonable assumption, since the binding force between CO$_2$ molecules is stronger than the binding force between CO$_2$ and water \cite{He2017co2diff}, and CO$_2$ molecules are prone to form clusters on ASW. Following He et al. (2017)\cite{He2017co2diff}, we describe the coverage of monomers and dimers using the following rate equations:
\begin{align}
  \frac{dC_1(t)}{dt} &= -2C_1(t)^2 k_D(T)\\
  \frac{dC_2(t)}{dt} &= C_1(t)^2 k_D(T)
\end{align} 
Note that both the left and the right sides of the equations have the unit of s$^{-1}$, since $C_1$ and $C_2$ are unitless.
The initial condition of the rate equation is $C_1(0)=0.075$ and $C_2(0)=0$. Solving the rate equation analytically, we have: 
\begin{equation}
  C_1(t)=\frac{1}{2k_D(T)t+\frac{1}{C_1(0)}}=\frac{1}{2k_D(T)t+13.33}
  \label{eq:c1_sol}
\end{equation}
$C_2(t)$ can be expressed as $0.5\times (C_1(0)-C_1(t))$. The next step is to relate $C_1(t)$ and $C_2(t)$ to the area of the 2330 and 2355~cm$^{-1}$ peaks. 
As discussed above, the 2330~cm$^{-1}$ peak area is not exactly proportional to $C_1(t)$ because $C_2(t)$ also contributes to it. However, it is reasonable to assume they have a linear relation, that is: 
\begin{equation}
  C_1(t) = \frac{a}{2k_D(T)t+13.33}+b  \label{eq:c1_sol_linear}
\end{equation}
where $a$ and $b$ are two free parameters. This is then used to fit the curves in the top panel of Figure~\ref{fig:bandArea}, and the fitting as well as the corresponding $k_D(T)$ value are shown in Figure~\ref{fig:Dfit}. 
The diffusion rate expression in Equation~\ref{eq:arr} can be rewritten as:
\begin{equation}
  \log(k_D(T))=\log(\nu_{\rm dif}) - \frac{E_{\rm dif}}{T} \label{eq:arr_modify}
\end{equation}
This shows that $\log(k_D(T))$ is a linear function of $1/T$ with a slope of $-E_{\rm dif}$. In Figure~\ref{fig:arrh} we make an Arrhenius-type plot. A linear fitting of it yields the slope and intercept, from which we obtain the diffusion energy barrier $E_{\rm dif}=1300\pm110$~K, and prefactor $\nu_{\rm dif}=10^{7.6\pm0.8}$~s$^{-1}$. In many other studies on diffusion, it is custom to report the Arrhenius expression of the diffusion constant (units of cm$^2$ s$^{-1}$) \cite{Tsong2005}: 
\begin{equation}
D=D_0 \exp(\frac{-E_{\rm dif}}{k_{\rm B}T})
\end{equation} 
$D_0$ is often called the diffusivity:
\begin{equation}
  D_0 =  \exp(\frac{\Delta S_{\rm dif}}{k_{\rm B}T})\frac{\nu_{0} l^2}{4}=\frac{\nu_{\rm dif} l^2}{4};
\end{equation} 
where $\Delta S_{dif}$ is the change in entropy between the saddle point and the adsorption site, and $\nu_0$ is the {\underline {small}}-amplitude frequency ($\approx 10^{13}$ sec$^{-1}$) of oscillation of the particle in the adsorption well.
In this case, D$_0$=$10^{-8.0\pm0.8}$ cm$^2$ s$^{-1}$.

\section{Discussions}
\label{sec:discuss}
Molecule diffusion is crucial for the chemistry on/in interstellar ice. However, laboratory measurements of diffusion are lacking, limiting the accuracy of astrochemical modeling of grain surface processes. The current study builds on prior studies \cite{He2017co2diff, He2018diff} to measure the diffusion energy barrier and prefactor for CO$_2$ on np-ASW surface simultaneously. A whole set of isothermal experiments were carried out to quantify the diffusion rate at different temperatures, from which the energy barrier and the prefactor for diffusion were then obtained. 

\begin{figure}[ht]
  \centering
  \includegraphics[width=0.9\textwidth]{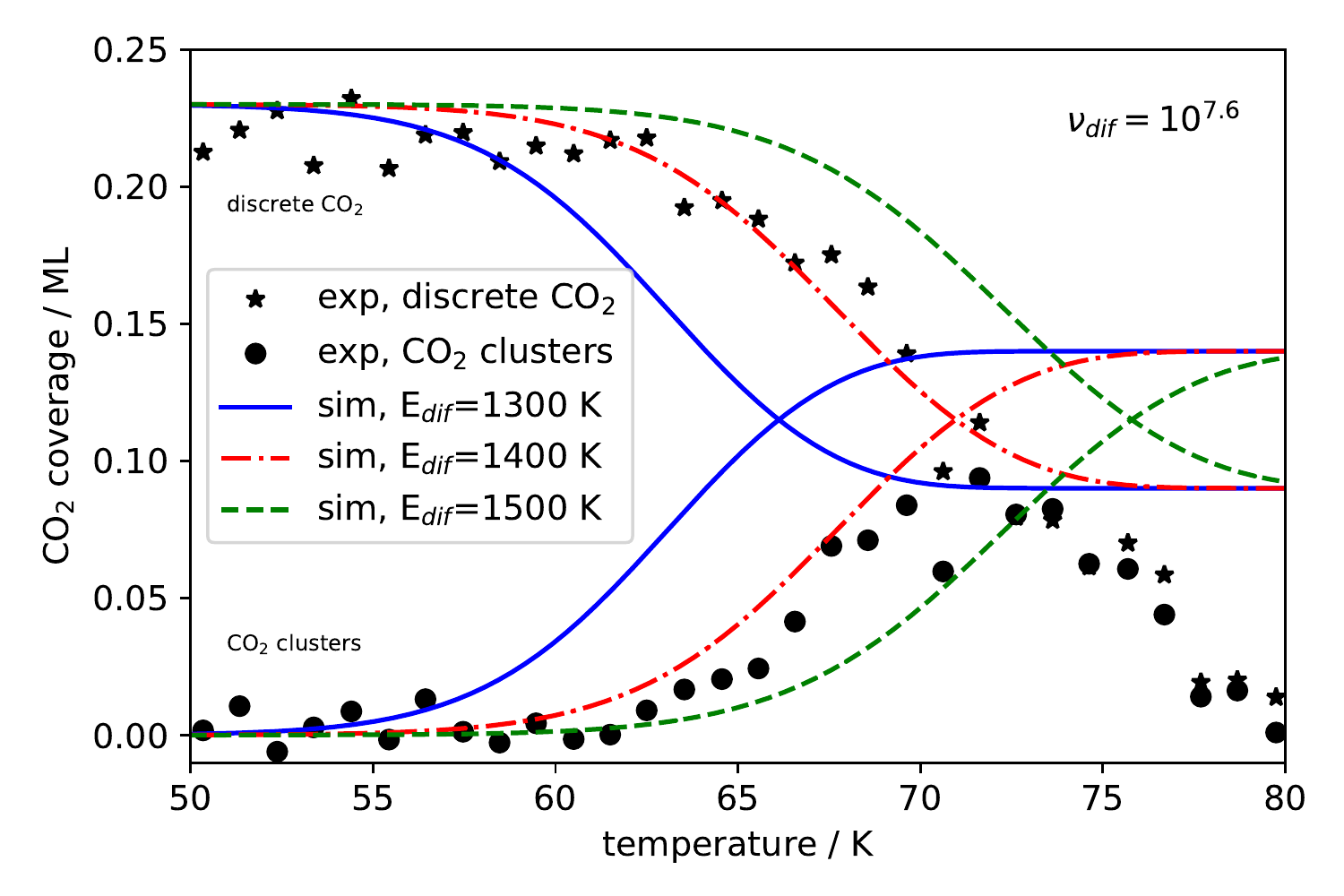}
  \caption{Recalculation of the diffusion energy barrier for the experiment in Fig.~10 of He17. The prefactor value $10^{12}$~s$^{-1}$ is replaced by $10^{7.6}$~s$^{-1}$. See He17 for more details. } \label{fig:he17fig10}
\end{figure}

It is worthwhile to compare this study to prior experimental studies. He17 deposited a small coverage of CO$_2$ on np-ASW and warmed up the ice slowly to monitor the shifting of infrared absorption peaks with increasing temperature. Because it was only a single measurement, it was not possible to derive the diffusion energy barrier and the prefactor simultaneously. It was assumed that the diffusion and desorption prefactor were the same, $\nu_{\rm dif} = \nu_{\rm des} = 10^{12}$~s$^{-1}$, and based on a rate equation model, the $E_{\rm dif}$ value was calculated to be $2150\pm50$~K, which is about 0.96 times the desorption energy of CO$_2$ from np-ASW surface. The main drawback of He17 is the assumption that the diffusion and desorption prefactors are equivalent, which is typically not the case, as demonstrated by He18. Here we recalculate the $E_{\rm dif}$ value for the experiment that is shown in Fig.~10 of He17, but replacing the $\nu_{\rm dif}$ value $10^{12}$~s$^{-1}$ by $10^{7.6}$~s$^{-1}$ obtained from the present study. Figure~\ref{fig:he17fig10} shows the equivalent of Fig.~10 in He17. By lowering the diffusion prefactor, the diffusion energy barrier is reduced to $1400\pm100$~K, which is in agreement with the value obtained by the present study $1300\pm110$~K within error allowance.

In the present study, we carried out isothermal experiments, similar to what He18 did, and indeed obtained a diffusion prefactor much lower than desorption, in agreement with He18. Based on prior studies on diffusion (see a review in \cite{Wang2002}), it is not uncommon to see different diffusion prefactors, either higher or lower than the one for desorption. This present study, along with He18, makes a strong case that in astrochemical modeling one should make a distinction between diffusion prefactor $\nu_{\rm dif}$ and desorption prefactor $\nu_{\rm des}$. Based on these two studies, we suggest using a $\nu_{\rm dif}$ value about 3--5 orders of magnitude lower than $\nu_{\rm des}$ in modeling. However, we acknowledge that this is only based on the measurement of a few volatile molecules on the ASW surface and that exceptions may be possible for other molecules. More diffusion measurements are required to confirm it.

Compared to He18, the Arrhenius fitting in Figure~\ref{fig:arrh} has a much larger error, and correspondingly, the uncertainty of diffusion energy barrier and prefactor values are larger. At least the following factors contribute to the larger uncertainty:
\begin{enumerate}
  \item The peak profile of CO$_2$ is more complicated than the dOH profile, as in He18. One cannot simply attribute the 2330~cm$^{-1}$ peak exclusively to CO$_2$ monomers. In our model, we had to simplify the problem by assuming a linear dependence between the coverage of monomers and the 2330~cm$^{-1}$ peak area. This could be one important source of uncertainty. 
  \item We assumed that there is only a single diffusion energy barrier that governs the formation of clusters. In practice, we cannot exclude the possibility that more than one activation energy is contributing. 
  \item A Gaussian function does not describe the 2355~cm$^{-1}$ component exactly. It is clear in Figure~\ref{fig:fitting} that the fitting of the 2355~cm$^{-1}$ peak is not ideal. 
  \item Ignoring clusters larger than dimers may also induce some uncertainty.
 \end{enumerate}  

A prior experimental determination of CO$_2$ on ASW surface was also done by Kouchi et al. (2020)\cite{Kouchi2020}, who reported an energy value of 1500 K, which is slightly higher than what we found in the present study. This difference could be due to the uncertainty in the prefactor that is assumed in Kouchi et al. They were unable to determine the prefactor value simultaneously with the energy value. It would be interesting to repeat the experiments in Kouchi et al. at different temperatures and derive the values of both the prefactor and the energy simultaneously. Karssemeijer \& Cuppen (2014)\cite{Karssemeijer2014} did molecular dynamics modeling calculation of CO$_2$ diffusion on water ice, and found the diffusion energy value to be $\sim1470$~K on a disordered water surface, which is close to our measured value. 

In the present study, we only presented the experiments on np-ASW. Although the porosity of the ice mantle on interstellar grains is still under debate, it is probably the case that it is neither highly porous nor completely compact. It is still relevant to consider the diffusion on the surface of both p-ASW and np-ASW. We attempted to run experiments similar to He18, i.e., tracing the diffusion of CO$_2$ into p-ASW by monitoring the infrared spectra. Unfortunately, the diffusion of CO$_2$ on p-ASW surface is not efficient enough to allow entry into the pores. CO$_2$ primarily remains  in the top layers as ``pure'' CO$_2$. The results are not shown here. Nonetheless, the results on np-ASW may apply to p-ASW as well. Previous experimental studies \cite{He2019asw, Zubkov2007} have suggested a similarity between the surface of a np-ASW and the pore surface of a p-ASW that is annealed to 60~K or above in terms of binding energy distribution and the fraction of dangling OH groups. The diffusion energy barrier and prefactor measured from np-ASW are likely to apply to the pore surface of p-ASW. We suggest that, in general, more experimental work is needed to measure diffusion on interstellar dust grain analogs.  

\section*{Acknowledgement}
We dedicate this paper to the memory of Dieter Gerlich, who was a good friend and colleague of some of us. We acknowledge the support from the European Research Council under the Horizon 2020 Framework Program via the ERC Advanced Grant Origins 83 24 28, and from NSF Astronomy \& Astrophysics Research Grant \#1615897. 
\bibliographystyle{tfo}
\bibliography{ref}

\end{document}